\documentstyle[epsfig]{mn2e}

\begin{document}
\title[Why are some galaxy disks extremely thin?]
{Why are some galaxy disks extremely thin?     }
\author[A. Banerjee and C.J. Jog ]
       {Arunima Banerjee$^{1}$\thanks{E-mail : arunima@ncra.tifr.res.in}, and
        Chanda J. Jog$^{2}$\thanks{E-mail : cjjog@physics.iisc.ernet.in}\\
$^1$  National Centre for Radio Astrophysics, TIFR, Pune 411007, India \\
$^2$   Department of Physics,
Indian Institute of Science, Bangalore 560012, India \\ 
}

\maketitle

\begin{abstract} 

Some low surface brightness galaxies are known to have extremely thin stellar disks with the vertical to
planar axes ratio 0.1 or less, often referred to as superthin disks.
Although their existence is known for over three decades, the physical origin 
for the thin distribution is not understood.
We model the stellar thickness 
for a two-component (gravitationally coupled stars and gas) disk embedded in a dark matter halo,
for a superthin galaxy UGC 7321 which has a dense, compact halo, and compare with a typical dwarf galaxy HoII which has a
non-compact halo.
We show that while the presence of gas does constrain the disk thickness, it is the compact dark matter halo which
plays the decisive role in determining the superthin disk distribution in low-mass disks.
Thus the compact dark matter halo significantly affects the  disk structure
and this could be important for the early evolution of galaxies.

\end{abstract}


\section{Introduction} 

The stellar disk in a typical spiral galaxy is thin, as seen from the image of an edge-on
galaxy like NGC 4565, and contains most of the visible mass in a late-type galaxy. 
There is a particular class of galaxies where the stellar disk is extremely thin, the so-called superthin (Goad \& Roberts
1981), or flat (Karachentsev 1989) galaxies,
where the vertical to radial axes ratio of the galaxy as projected on the sky is 0.1 or less (Matthews et al. 1999), 
also see Kautsch (2009) 
for a recent review.

There is no clear physical understanding yet of why the stellar disks
in these galaxies are thin. This is of interest for understanding the structure and dynamics of these galaxies. 
In addition, the superthin galaxies have become interesting recently for two reasons. First, all the superthin galaxies are  low surface brightness (LSB) galaxies though the reverse is not true. Thus understanding why these are so thin may help shed light on the
overall evolution of LSB galaxies. The LSBs are common, are highly gas-rich yet do not appear to have had a significant star formation, and their origin and dynamics are not understood well (e.g., de Blok 2006).  
Second, galaxy interactions are now known to be common and can thicken the stellar disks (Walker et al. 1996, Velazquez \& White 1999, Qu et al. 2011). Irrespective of what gives rise to the thick disks, most large galaxies are now known to contain
a thick disk (Gilmore \& Reid 1983, Dalcanton \& Bernstein 2002). In view of this, the existence of galaxies with a superthin stellar disk is even more puzzling.

The vertical thickness of a galactic stellar disk is determined by the balance between the vertical self-gravitational force 
and pressure (e.g., Rohlfs 1977). 
The inclusion of gas, despite its low mass fraction, has a significant constraining effect in the determination of the stellar thickness. This is because the gas being a low-dispersion component is located closer to the mid-plane (Narayan \& Jog 2002b, Banerjee \& Jog 2007).
The LSB galaxies are known to be dark-matter rich from the innermost radii of the galactic disk (McGaugh \& de Blok 1998, Banerjee et al. 2010). 
Thus the halo is likely to play an important role in reducing the stellar thickness in these, since the surface density of 
dark matter within a constant thickness
near the mid-plane dominates the disk surface density from the inner regions (see Fig. 3, Banerjee et al. 2010).

In this paper, we study the stellar thickness for a two-component (stars plus gas) disk in a dark matter halo.  
We study the prototypical superthin galaxy UGC 7321 for which the various input parameters for the stellar and gas components have been determined from observations
(Matthews et al. 1999, Matthews 2000, Uson \& Matthews 2003, Matthews \& Wood 2003). We obtain the stellar scale height at different radii for UGC 7321 using the two-component model of
gravitationally-coupled stars and gas in the field of a dark matter halo as developed in 
Banerjee et al. (2010), 
and explore the parameter space around the observed values to get an idea of the variation in the 
calculated scale height. We show that while the presence of gas constrains the stellar disk thickness, it is the dense, compact dark matter halo 
which plays a decisive role in determining whether the disk is thin.

Our earlier studies where the vertical thickness of the neutral 
\mbox{H\,{\sc i}} gas was used to constrain the shape and density profile of the dark matter halo  
modeled as pseudo-isothermal spheres 
revealed an interesting feature, namely that large ordinary spirals like the Galaxy and M31 have non-compact halos while
the superthin galaxy UGC 7321 has a compact halo (\S 2.2).
This has prompted us to investigate here if in general  a dense, compact halo 
plays a crucial role in determining the superthin distribution in a stellar disk. 

The paper has been organized as follows. In \S 2, we present the numerical calculations, definitions of the terms and input parameters used, 
in \S 3 the results, followed by discussion and conclusions in \S 4 and \S 5 respectively.

\section{Numerical Calculations \& Definition of Terms Used}

\subsection{Numerical Calculations}

We use the two-component model of gravitationally-coupled stars and \mbox{H\,{\sc i}} gas in a galactic disk in the  field of a pseudo-isothermal dark matter halo. 
In this model, the stars and gas are assumed to be present in the form of concentric, thin, axisymmetric disks embedded within each other, and are in a hydrostatic equilibrium in the $z$ direction (Narayan \& Jog 2002b, Banerjee \& Jog 2008).
We use the galactic cylindrical co-ordinates 
(R, $\phi$ and z).

The joint Poisson equation for the above two-component disk and halo is given by:

\begin{equation}
\frac{1}{R}\frac{\partial}{\partial R}(R\frac{\partial \Phi_{total}}{\partial R}) +  \frac{\partial^2\Phi_{total}}{\partial z^2} = 4\pi G(\sum_{i=1}^{2} \rho_i + \rho_{h})  
\end{equation}

\noindent where $\Phi_{total}$ is the net potential due to the stars, the \mbox{H\,{\sc i}} gas and the dark matter halo; $\rho_i$ with $i = 1$ to 2 denotes the mass volume density for each of the disk components (stars, and HI gas),
 while $\rho_h$ denotes the same for the pseudo-isothermal halo. The halo density profile is given by (Binney \& Tremaine 1987):

\begin{equation}
\rho_h(R,z) = \frac{\rho_0}{1+\frac{R^{2} + z^{2}}{R_c^{2}}} 
\end{equation}

\noindent where $\rho_0$ is the central density and $R_c$ is the core radius of the halo.

The equation of hydrostatic equilibrium for the $i^{th}$ disk component in the $z$ direction is given by (Rohlfs 1977):

\begin{equation}
\frac{\partial }{\partial z}(\rho_{i}<(v_{z}^{2})_{i}>) + \rho_{i}\frac{\partial \Phi_{total}}{\partial z} = 0
\end{equation}

On combining the above two equations, we get the equation for the vertical equilibrium for each component of the disk to be:
\begin{equation}
{\langle v_{z}^{2} \rangle}_{i} \frac{\partial}{\partial z}\left[\frac{1}{\rho_{i}}\frac{\partial \rho_{i}}{\partial z}\right] = -4\pi G(\sum_{i=1}^{2} \rho_i + \rho_{h}) + 
\frac{1}{R}\frac{\partial }{\partial R}{({{v_{rot}}^{2}(R))}_{obs}}
\end{equation}
\noindent Here 
$<v_{z}^{2}>_{i}$ is the mean square velocity of the $i^{th}$ disk component, and $ { {(v_{rot})}_{R} } $ is the observed rotational velocity at any R. The last term is the radial part of the Laplacian. For a flat rotation curve
this is identically equal to zero at the mid-plane, and for a nearly flat curve it changes the vertical scale heights less than 1\% and hence can be ignored (Narayan et al. 2005). 
However, for a region of rising rotation curve as in dwarf galaxies, including this term can increase the scale heights 
by about 15 \% (Banerjee et al. 2011), so we include it here for the calculations for HoII (Section 3.2). 
For UGC 7321, we have not included the radial term in the calculations to be consistent with our earlier work (Banerjee et al. 2010), the results of which such as $\rho_0$ and $R_c$ have been used as input parameters in the
 current paper. The error in the calculated scale height is estimated to be $\sim$ 22 $\%$, thus even if this term were to be included, we would still get a superthin distribution for UGC 7321.

 Therefore, we have two
coupled non-linear, second-order ordinary differential equations in the variables $\rho_{stars}$ and $\rho_{gas}$. These are solved
numerically using the fourth-order Runge-Kutta method of integration in an iterative fashion, using suitable initial conditions 
(for details see Narayan \& Jog 2002b, Banerjee \& Jog 2008).
The stellar density is obtained as a function of $z$ at each galactocentric radius R, 
and the half-width-at-half-maximum (HWHM) defines the vertical thickness at each R. 

\subsection{Definitions of Terms Used}

\noindent {\it A superthin galaxy} :
In this paper we use the ratio $z_0/R_D$ where $z_0$ is the vertical stellar scale height as in 
$\rho = \rho_0$ Sech$^2(z/z_0)$ and ``$R_D$'' is the exponential
stellar disk scale length as the quantitative index of the stellar disk thickness following the work of Reshetnikov et al. (2003)
and Bizyaev \& Kajsin (2004). 
At each $R$, $z_0$ is obtained from the corresponding HWHM using a simple conversion factor ($z_0$ = HWHM / 0.88).  
We refer to a galaxy as being superthin if the average value of $z_0/R_D$ within $R \leq 3 R_D$ is $\leq 0.1$

\noindent {\it Dense and compact Halo}: In this work, we define the dark matter halo as being ``dense'' if the central density $\rho_{0}$ is $\sim$ 
a few times 0.01 
M$\odot$pc$^{-3}$ or more. 
Ordinary large spirals like the Galaxy ($\rho_{0}$ $\sim$ 0.033 M$\odot$pc$^{-3}$; Narayan et al. 2005), 
M31 ($\rho_{0}$ $\sim$ 0.011 M$\odot$pc$^{-3}$; Banerjee \& Jog 2008), and the superthin galaxy UGC 7321 
($\rho_{0}$ $\sim$ 0.035 - 0.057 M$\odot$pc$^{-3}$; Banerjee et al. 2010) thus have dense halos. 
Similarly, we label the halo to 
be ``compact'' if the core radius $R_c \leq 2 R_D$. 
The high surface brightness (HSB) galaxies are found to have non-compact or extended halos with 
$R_c \geq 2 R_D$ (Gentile et al. 2004). This is confirmed for our Galaxy where $R_c/ R_D \sim 2.8$ (Narayan et al. 2005),
and M31 where $R_c/ R_D \sim 3.9$ (Banerjee \& Jog 2008).
The superthin 
galaxy UGC 7321, on the other hand, has been shown to have a compact dark matter halo with $R_c /R_D \sim 1.2-1.4 $ (Banerjee et al. 2010). 
In contrast, the dwarf irregular galaxy such as HoII studied in \S 3.2 here, has less dense and extended halo, with  $\rho_{0}$ = 0.0009 M$\odot$pc$^{-3}$ and $R_c/ R_D \sim 8.3 $.

\subsection{Input Parameters Used}

We choose UGC 7321 as a prototypical example for our study as it is one of the very few superthin galaxies 
reasonably well studied observationally (\S 1), and
all the required input parameters for our model are available, as indicated in Table 1.
The choice of stellar vertical velocity used is discussed next in some detail, since the stellar thickness depends crucially on this. 
The vertical stellar dispersion was measured by Matthews (2000) indirectly
by assuming pressure equilibrium along the vertical direction for a stellar component alone.  From the observed
surface density (obtained from the luminosity and a M/L ratio) and the scale height for stars,
and using eq. (7) in Matthews (2000), the central value of the vertical velocity
dispersion is obtained to be 14.3 km s$^{-1}$ (Banerjee et al. 2010). Here for consistency, we use the same value 
since we use the values of the halo parameters derived in 
Banerjee et al (2010) which were based on this dispersion value.
We note that even the addition of gas and dark matter densities
(from Fig. 3, Banerjee et al. 2010)  will increase the central stellar dispersion
by only $\sim 20 \%$ to 17.0 km s$^{-1}$. We check that the resulting 
error in the calculated scale height is then 21 \% and the calculated disk 
continues to be a superthin one.

We assume that the velocity dispersion falls off exponentially with a scale length, $R_v$, equal to $2R_D$.
The measurement of disk scale length in an edge-on galaxy could be affected by the
dust extinction which could be uneven or patchy.
We have taken $R_D = 2.1$ as measured 
in R-band (Matthews 2000) which is therefore not too affected by the dust 
extinction. The K-band value is a better indicator of the true 
underlying stellar mass distribution, this was measured using the 2MASS data 
by Bizyaev \& Mitronova (2002) who get a value that is not significantly different.

\begin{center}
\begin{table*}
{\small
\hfill{}
\caption{Input parameters of UGC 7321$^{1}$ and HoII$^{2}$}
\begin{tabular}{|l|l|c|c|c|c|c|c|c|c|}
\hline
\hline
Galaxy & &${\Sigma_*(0)}$ & ${\Sigma_{Gas}}^{3}$ & ${R_{D}}$ & ${\sigma_z}^{HI}$ &${\sigma_z}^{stars}(0)$ & ${\rho_{0}}$ & ${R_{c}}$ & $R_c /R_D$  \\
      & & (M$_{\odot}$\,pc$^{-2}$) & (M$_{\odot}$\,pc$^{-2}$) & (kpc) & (km\,s$^{-1}$) & (km\,s$^{-1}$) & (M$_{\odot}$\,pc$^{-3}$) & (kpc)     \\
&&&&&&&&&\\
\hline
UGC 7321 & & 50.2  & 4.1 & 2.1 & 9 & 14.3 &0.057 & 2.5 & 1.38   \\
HoII     & & 27.8 & 9.8 & 1.2 &7 &11.1 &0.0009 & 10.0 & 8.33   \\
\hline
\small $^{1}$ Banerjee et al. 2010 \\
\small $^{2}$ Banerjee et al. 2011  \\
\small $^{3}$ For R $\leq$ 3$R_D$
%end{center}
\end{tabular}}
\hfill{}
\label{tb:tablename}
\end{table*}
\end{center}

\section{Results}

\subsection{UGC 7321: A galaxy with a dense, compact halo}

We now study the relative importance of the gas surface density with respect to a dense, compact dark matter halo in determining the vertical 
thickness of a stellar disk.
Using our two-component model of a galaxy (see \S 2.1), we
obtain the vertical scale height of the stars at each galactocentric radius R for UGC 7321, a galaxy with a superthin stellar disk.

In Figure 1, we plot the ratio
 ${z_0}/{R_D}$ versus R for the stellar component in UGC 7321 for four different cases as indicated. 
These are the two-component model with the stars responding only to their own self-gravity 
i.e model excluding self-gravity of the gas and the gravitational field of the dark matter halo, 
model excluding only the gravitational field of dark matter, 
model excluding only the self-gravity of the gas, 
and lastly, the complete model including all the components i.e the self-gravity of the stars, the gas and the field of the 
dark matter halo. It is clear from the figure that the stellar disk of UGC 7321 qualifies as a superthin case by our definition (${z_0}/{R_D} \leq 0.1$) only  
when the gravitational field of the dark matter halo is taken into account. 
This clearly illustrates the fact that although the self-gravity of the gas plays a role in reducing the vertical thickness of the stars 
as compared to that of the stars-alone case,  
it is the gravitational force due to the dark matter halo that is principally responsible in causing the
stellar disk to be superthin.

\begin{figure}
\centering
\includegraphics[width=2.9in,height=2.4in]{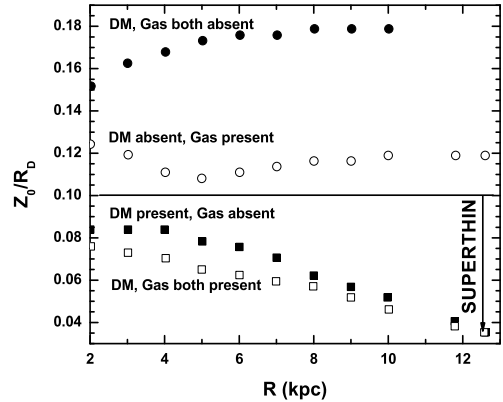} 
\caption{Plot of the ratio $z_0/R_D$ (where $z_0$ is the stellar vertical scale height and $R_D$ the exponential disk scale length) with 
galactocentric radius R for UGC 7321 for different realizations of the two-component galaxy model. 
From top to bottom, the curves correspond to the two-component model with both gas and dark matter absent, with gas present but dark matter absent, 
with gas absent but dark matter present,
and with both gas and dark matter present respectively. Thus the ratio $z_0/R_D$ is found to  
depend both on the
gas and the dark matter halo components. For instance, at R = 2 kpc, inclusion of
gas (dark matter) reduces this ratio by $\sim$ 18$\%$ (45$\%$) to the value 0.124 kpc (0.084 kpc) as compared to the stars-alone case (0.152 kpc). 
However, this figure clearly shows that it is the dense, compact dark matter halo and not the gas-rich disk, that is responsible for the superthin stellar disk observed in 
UGC 7321.}
\end{figure}

In contrast, in the Galaxy the stellar disk is substantially more massive than in UGC 7321, hence the disk gravity dominates the vertical self-gravitational force, and  
the halo plays a minor role in determining the vertical stellar thickness in the inner Galaxy as was shown by Narayan \& Jog (2002b). 
In the inner Galaxy, the rotation curve is mainly set by the disk, while in UGC 7321 the surface density of stars is much lower 
and hence stars contribute little to the radial gravitational force. Hence a dense and compact dark matter halo is also to explain the observed
rotation curve in UGC 7321 as well.

In Figure 2, we study the importance of the compactness of the dark matter halo in
determining the vertical thickness of the stellar disk in UGC 7321. Using our two-component model of gravitationally-coupled stars 
and gas in the field of a dark matter halo, but with the self-gravity of the gas set to zero, we obtain 
 the average of $z_0/R_D$ (averaged over R $\leq$ 3$R_D$) for different values of the dark matter core radius $R_c$, for
a given asymptotic value of the rotational velocity ${v_{rot}}(R=\infty)$ = 110 km s$^{-1}$ as set by the 
rotation curve from the relation $ v_{rot}^2 = 4 \pi G \rho_0 {R_c}^2 $. This, in turn, makes the dark matter core density $\rho_0$ a function of the core radius $R_c$. 
Also, we choose a feasible range of trial values for $R_c$ (0.5$R_D$ $\leq$ $R_c$ $\leq$ 5$R_D$) as indicated by earlier studies.
The plot clearly brings out that a superthin stellar disk (average $z_0/R_D$ $\leq$ 0.1) results 
only when $R_c \leq 2R_D$ i.e when the dark matter halo is compact. We also study the relative importance of the gas 
surface density as compared to the compactness of the dark matter halo in this figure.
Therefore also overlaid on this plot are horizontal lines 
indicating the average value of $z_0/R_D$ for different values of the gas to stellar surface density fraction, 
 with the gravitational field due to the dark matter 
set to zero. This shows that in order to obtain a superthin stellar disk the gas fraction should be more than the 
the observed gas surface density in UGC 7321 in the region of interest. 
Therefore this figure again highlights the fact that a compact dark matter halo is necessary
to obtain the superthin stellar disk in UGC 7321. Thus, only a high value of the relative mass of the dark matter halo can not be the reason behind the superthin stellar disk of a galaxy as claimed by Zasov et al. (2002). Instead the actual radial distribution of the dark matter is important.

\begin{figure}
\centering
\includegraphics[width=2.9in,height=2.4in]{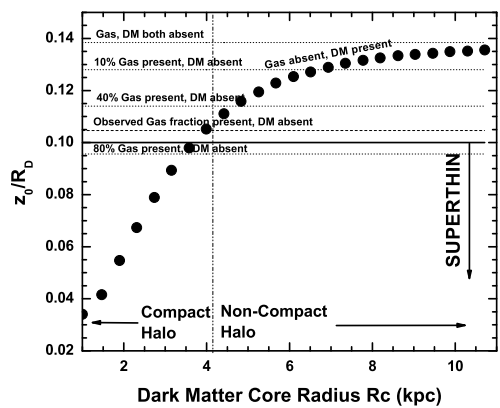} 
\caption{Plot of the average value of the ratio $z_0/R_D$ (where $z_0$ is the stellar vertical scale height and $R_D$ the exponential disk scale length) as a 
function of the dark matter core radius $R_c$, here $R_D = 2.1 $ kpc as in UGC 7321.
From the plot it is clear that the stellar disk becomes superthin only for compact dark matter halos i.e $R_c \leq 2R_D$. Also overlaid on this figure
are horizontal cuts indicating the average $z_0/R_D$ for cases with different gas surface density fractions but with that of the dark matter component set to zero. 
It is evident that in order to render the stellar disk superthin through a gas-rich disk only, the fraction of the gas surface density needs to be more ($\sim$ 70$\%$ of that of the stars) 
than the observed value in UGC 7321 ($\sim$ 47$\%$ within the region of interest i.e.,R $\leq$ 3$R_D$). This again confirms that a dense, compact dark matter halo is necessary for a superthin stellar disk.}
\end{figure}

Finally, we repeat the same calculations as above but with a central stellar surface density 
half of its original value to study the
dependence of the vertical thickness of the stellar disk on the stellar surface density, as discussed in
Bizyaev \& Kajsin (2004). Most interestingly, even the case with the gas surface density equal to the stellar surface density, 
 fails to result in  
a thin stellar disk. In contrast,  a highly compact halo with $R_c \leq 1.5 R_D$ 
can render the stellar disk superthin even without any contribution from the gas component. Therefore, for low-mass galactic disks, 
a dense and compact dark matter halo plays the pivotal role in 
making the stellar disk superthin.  

It may be appear that a very thin stellar disk could simply be explained by a low vertical dispersion and hence lower 
pressure support. There is indeed
 reason to believe that the stellar dispersion could be lower in the superthin galaxies
since the various dynamical processes for heating the stellar dispersion are not effective (Section 4).
However, from the results in Figures 1 and 2, we can argue that the low 
dispersion alone cannot be responsible for the superthin nature of UGC 7321.
In absence of the dark matter, the disk is thick enough (with an average 
$z_0/R_D$ well above 0.1) so that it will not be considered to be a 
"superthin" galaxy (See Fig. 1). Further, Figure 2 shows that for the given observed 
stellar velocity dispersion, the disk shows a "superthin" distribution 
only when the dark matter halo is compact (with $R_c/ R_D < 2$).  
Thus a low stellar dispersion cannot be the main reason for the superthin distribution.

\subsection{Comparison with HoII, a dwarf galaxy with a non-compact halo}

In order to illustrate further the relative importance of a dense, compact dark matter halo and a gas-rich disk
 in determining the stellar disk thickness in realistic cases, we next apply our two-component model of the galaxy 
to obtain the average $z_0/R_D$ 
for the stars in the dwarf irregular galaxy HoII chosen from the THINGS sample. 
HoII is gas-rich with the gas surface density dominating the disk dynamics on an equal footing with the 
stars. However the dark matter halo is neither dense (${\rho}_{0}$ = 0.0009 $M_{\odot} pc^{-3}$) and nor compact ($R_c/R_D \sim 8.3$). 
The input parameters required for our calculations are presented in Table 1. The central stellar velocity
dispersion has been measured to be 11.1 km s$^{-1}$ (Banerjee et al. 2011), using the approach as in Leroy et al. (2008). The dispersion is assumed to fall off exponentially with radius with a scale length equal to 2$R_D$.
In Table 2, we compare the resulting average $z_0/R_D$
for HoII with that of UGC 7321 for different cases. 
Interestingly, the dense, compact halo of UGC 7321 can reduce its average $z_0/R_D$ by
50$\%$ compared to the value of the stars-alone case. The non-compact halo of HoII, on the other hand,  
can reduce the stars-alone value by 10$\%$ only. Inclusion of gas to the stars-alone case, on the other hand,
reduces the same by 18$\%$ and 56$\%$ respectively. Therefore although the highly gas-rich disk of HoII is as 
effective as the dense, compact halo in reducing the disk thickness, it cannot, however, make the 
stellar disk superthin. This result corroborates our observation in \S 3.1 that in a low-mass disk, only a dense, 
compact dark matter halo can render the disk superthin.

%\begin{center}
\begin{table}
%\small
%\hfill{}
%\caption{Average resulting ratio of the vertical scale height to the exponential disk scale length $ {z_0}/{R_D}$ for UGC7321 and HoII}
\caption{Average $ {z_0}/{R_D}$ for UGC7321 and HoII}
\label{table2}
%\centering
\begin{tabular}{|l|c|c|}
\hline
Model & UGC7321 & HoII  \\
\hline
      &         &         \\
DM, \& Gas both absent & 0.138 & 0.374 \\
DM absent, Gas present & 0.113 & 0.166 \\
DM present, Gas absent & 0.070 & 0.335 \\
DM present, Gas present & 0.068 & 0.132\\
\hline
\end{tabular}
%\hfill{}
%\label{tb:tablename}
\end{table}
%\end{center}

Before concluding this section, we note that although each of the above two galaxies constitutes a typical representative of their 
respective galaxy class, namely UGC 7321 for the superthin LSBs and HoII for the dwarf irregulars, the values of
the physical parameters would vary from galaxy to galaxy even within the same class of galaxy. Nevertheless varying 
the relevant input parameters around the realistic values seen in these prototypes and studying the effect on the stellar disk thickness 
helps us to follow the
underlying trend, thereby providing useful pointers for future research.

\subsection{Physical basis of a compact halo leading to a superthin stellar disk}

For a dark matter-dominated galactic disk, the vertical thickness of the stars at any radius R is regulated by the dark matter mass density
$\rho_h$ close to the midplane ($z = 0$), or rather, by the steepness of its density profile. This is because the stars effectively 
respond as test
 particles
to the dominant mass component of the disk, which is the dark matter halo in this case. Now the steepness of the density profile in the vertical 
direction is given by its first derivative with respect to $z$ i.e ${d\rho_h}/{dz}$ which is zero at
$z = 0$ for the pseudo-isothermal dark matter halo. So we use the second-derivative to quantify the steepness of the density profile. 
Simple calculations show that 
for a given value of the asymptotic rotational velocity, the second-derivative of the dark matter density $\rho_h$ at any R is given by
${d^2\rho_h}/{dz^2} \sim {1}/{ {({R_c}^2 + R^2)}^2 }$. Therefore a small value of $R_c$ 
will result in a high value of ${d^2\rho_h}/{dz^2}$. In other words, a compact dark matter halo will result in a 
steep vertical density profile
of the dark matter halo, and therefore, of the stars, which is equivalent to a thin stellar disk.

\section{Discussion}

\noindent {\it 1. Results for a slower fall-off in stellar dispersion} :
In the calculations above, it was assumed that the vertical velocity dispersion 
falls exponentially with radius with a scale length, $R_v$, that is twice the 
scale length of the disk exponential surface brightness, $R_D$ (or, $R_v/R_D = 2$). 
This standard relation was obtained 
for a single-component disk of constant thickness (van der Kruit \& Searle 1981).
In an earlier work (Narayan \& Jog 2002a) it was shown that the 
observations of luminosity profiles when looked at closely do not imply rigorously constant 
thickness, indicate a moderate flaring of stellar thickness, by a factor of few within 
the optical radius. 
This issue was studied for a multi-component disk in a dark 
matter halo, and comparing the model results with observations of luminosity profiles, it was shown that 
the ratio $R_v/R_D$  lies in a range of 2-3 (Narayan \& Jog 2002a).

Choosing a typical value of 2.5 in this range, we next repeat the 
calculation - see Figure 3 for the results.
We find that 
for this higher ratio, the fall off in stellar velocity dispersion is slower, 
hence the resulting $z_0/R_D$ values are higher especially at larger radii as expected (see Figure 3). Nevertheless, the 
two main features of Figure 1 still remain. First, the disk is still superthin 
(with the average $z_0/R_D <$ 0.1 for radii $< 3 R_D$). Second, the main 
deciding factor for the superthin nature of the disk is the compact dark matter halo 
(and not the gas), as can be seen by comparing the two curves that include the 
dark matter to those without it.
Thus our main result from this paper, namely that a compact halo is the principal 
determining factor that determines the disk thinness, remains valid even for a 
higher value of $R_v/R_D$ that was indicated by an earlier study (Narayan \& Jog 2002a).

\begin{figure}
\centering
\includegraphics[width=2.9in,height=2.4in]{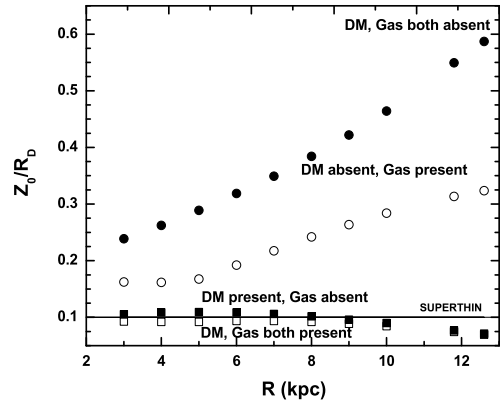} 
\caption{Plot of the ratio $z_0/R_D$ 
versus  galactocentric radius R for UGC 7321 for different realizations of the two-component galaxy model (as in Figure 1), but with a more gradual fall-off with radius in the stellar vertical velocity dispersion (as given by $R_v/R_D$ = 2.5). As expected, the scale height values are higher here due to the higher pressure support from the stellar component.  Nevertheless, the 
two main features of Figure 1 still remain: namely, the disk is still superthin 
(with the average $z_0/R_D < 0.1$ for radii within $< 3 R_D$); and the main 
deciding factor whether the disk is thin is the compact dark matter halo 
(and not the gas), as can be seen by comparing the two curves that include the 
dark matter to those without it.}
\end{figure}

\noindent {\it 2. Evolution of disk thickness} : Galaxy interactions are now known to be common, and these tend to increase
the thickness of stellar disk with time (\S 1).
Interestingly, the LSBs are known to lie in a region of low galaxy number density 
(Karachentsev 1999). The analysis of SDSS data shows these to lie along the edges of voids (Rosenbaum et al. 2009). This lack of galaxy interactions could explain why superthin galaxies retain their very thin stellar disk distribution
over time.
Bars and spiral features can thicken a stellar disk (Saha et al. 2010), but UGC 7321 has no detectable bar (Pohlen et al. 2004), and LSBs generally lack spiral features (de Blok 2006). So this mode of disk thickening does not apply in this case. Also, bending instabilities that could lead to disk thickening may also not be effective in presence of a dominant dark matter halo (Zasov et al. 1991). Thus the various dynamical processes that could cause a disk thickening
are ineffective for the LSB galaxies. Hence, if the stellar disk is superthin to begin with, that property remains unchanged with time.

\noindent {\it 2. Origin of a dense, compact dark matter halo} : The main underlying question from this work is: what decides whether the dark matter halo is dense
and compact? Does the presence of a disk in it affect its internal evolution via angular
momentum transfer? It is still not clear why two disks with otherwise similar properties 
(in terms of the stellar surface density and gas fraction) but with different halo properties give a superthin stellar disk 
(as in  UGC 7321) or
a puffed up disk (as in HoII). We hope that the result in present paper will trigger further research 
involving high resolution simulations of galaxy evolution that could address this issue.

\section{Conclusions}

We attribute the superthin nature of the stellar disk seen in some low surface brightness galaxies to the presence of a dense and {\it compact} dark matter halo.
We obtain the stellar disk thickness for gravitationally coupled stars and gas in the presence of a 
dark matter halo. By studying the parameter space around the observed values for the superthin galaxy UGC 7321, we show that while the inclusion of gas in the problem does constrain the stellar disk thickness, 
 the superthin distribution is mainly attributed to a dense, compact dark matter halo which dominates the disk from the inner radii of the galaxy. 
Thus a compact dark matter halo plays a crucial role in determination of the structure and dynamics of superthin galaxies. The question that needs to be addressed is what gives rise to 
such dense and compact dark matter halos to begin with during the galaxy evolution phase.
Since superthin low surface galaxies constitute the prototypes of the high redshift galaxies in the local universe, our result may be
have important implications for galaxy formation and evolution models. 

\section*{Acknowledgment}

We thank Lynn Matthews, Aparna Maybhate, Paola Di Matteo and Rajaram Nityananda for useful discussion on 
related projects. 

\bigskip

\section*{References}

\medskip

\noindent  Banerjee, A., \& Jog, C. J. 2007, ApJ, 662, 335

\noindent Banerjee, A., \& Jog, C. J. 2008, ApJ, 685, 254 

\noindent Banerjee, A., Matthews, L. D., \& Jog, C. J. 2010, New~A, 15, 89 

\noindent Banerjee A., Jog, C. J., Brinks, E., \& Bagetakos, I. 2011, MNRAS, 415, 687 

\noindent Binney, J., \& Tremaine, S. 1987, Galactic Dynamics. Princeton Univ. Press, Princeton, NJ 

\noindent Bizyaev, D., \& Mitronova, S. 2002, A\&A, 389, 795

\noindent Bizyaev, D., \& Kajsin, S. 2004, ApJ, 613, 886

\noindent Dalcanton, J., \& Bernstein, R. A. 2002, AJ, 124, 1328

\noindent de Blok, W.J.G. 2006, in 
{\it Encyclopedia of Astronomy \& Astrophysics}, P. Murdin, Ed.,
(Bristol: IOP), p. 1

\noindent Gentile, G., Salucci, P., Klein, U., Vergani, D., \& Kalberla, P. 2004, MNRAS, 351, 903

\noindent Gilmore, G., \& Reid, N. 1983, MNRAS, 202, 1025

\noindent Goad, J.W., \& Roberts, M.S. 1981, ApJ, 250, 79 

\noindent Karachentsev, I. 1989, AJ, 97, 1566

\noindent Karachentsev, I. 1999, Astr.L., 25, No.5, 318

\noindent Kautsch, S.J. 2009, PASP, 121, 1297

\noindent Leroy, A. K., Walter, F., Brinks, E., Bigiel, F., de Blok, W. J. G., Madore, B., Thornley, M. D. 2008, AJ, 136, 2782 

\noindent Matthews, L. D., van Driel, W., \& Gallagher, J. S. 1999, AJ, 118, 2751

\noindent Matthews, L. D. 2000, AJ, 120, 1764

\noindent  Matthews, L. D., \& Wood, K. 2003, ApJ, 593, 721

\noindent  McGaugh, S.S., \& de Blok, W.J.G. 1998, ApJ, 499, 41

\noindent  Narayan, C. A., \& Jog, C. J. 2002a, A\&A, 390, L35

\noindent  Narayan, C. A., \& Jog, C. J. 2002b, A\&A, 394, 89

\noindent  Narayan, C. A., Saha, K., \& Jog, C. J. 2005, A\&A, 440, 523

\noindent  Pohlen, M., Balcells, M., \& Matthews, L.D. 2004, ASSL, 319, 791 

\noindent  Qu, Y., Di Matteo, P., Lehnert, M.D., \& Van Driel, W. 2011, A\&A, 530, A10

\noindent  Reshetnikov, V. P., Dettmar, R.-J., \& Combes, F. 2003, A\&A, 399, 879

\noindent  Rohlfs, K. 1977, Lectures on Density Wave Theory (Berlin: Springer-Verlag)

\noindent  Rosenbaum,S.D., Krusch, E., Bomans, D.J., \& Dettmar, R.-J. 2009, A\&A, 504, 807

\noindent  Saha, K., Tseng, Y.-H., \& Taam, R.E. 2010, ApJ, 721, 1878

\noindent  Uson, J.M.,\& Matthews, L. D. 2003, AJ, 125, 2455

\noindent  van der Kruit, P.C., \& Searle, L. 1981, A\&A, 95, 105

\noindent  Velazquez, H., \& White, S.D.M. 1999, MNRAS, 304, 254

\noindent  Walker, I.R., Mihos, J.C., \& Hernquist, L. 1996, ApJ, 460, 121 

\noindent  Zasov, A.V., Makarov, D.I., \& Mikailhova, E.A., 1991. Soviet Aston. Lett. 17, 374

\noindent  Zasov, A.V., Bizyaev, D.V., Makarov, D. I., \& Tyurina, N. V. 2002,
Astronomy Letters, 28, 527

\end{document}